\documentclass[lettersize,journal]{IEEEtran}
\usepackage[T1]{fontenc}

\usepackage{cite}
\usepackage{amsmath,amssymb,amsfonts}
\usepackage{algorithm}
\usepackage{stfloats}
\usepackage{graphicx}
\usepackage{textcomp}
\usepackage{algpseudocode}
 \usepackage{booktabs}

\usepackage{amsthm}
\usepackage{balance}
\usepackage{url}
\providecommand{\refname}{References}

\usepackage{multirow}
\usepackage{multicol}
\theoremstyle{plain}

\theoremstyle{definition}
\newtheorem*{assumption*}{Assumption}

\theoremstyle{remark}
\newtheorem*{remark*}{Remark}

\def\BibTeX{{\rm B\kern-.05em{\sc i\kern-.025em b}\kern-.08em
    T\kern-.1667em\lower.7ex\hbox{E}\kern-.125emX}}
\begin{document}
\bstctlcite{IEEEexample:BSTcontrol}

\title{Observable and Null-Space Electrophysiological Source Imaging with Diffusion Priors}

\author{Zirui Wen$^*$, Shihao Yang$^*$, Feng Liu
\thanks{Zirui Wen, Shihao Yang, and Feng Liu are with the
Department of Industrial and Systems Engineering,
Rutgers University, New Brunswick, NJ 08901, USA. $^*$ denotes co-first authors.
Research reported in this publication was supported by the National Institute of Neurological Disorders and Stroke of the National Institutes of Health, United States under Award Number R21NS135482 (PI: Liu). The content is solely the responsibility of the authors and does not necessarily represent the official views of the National Institutes of Health.}%
\thanks{Corresponding author: Feng Liu. E-mail: feng.liu@rutgers.edu.}

}

\maketitle

\begin{abstract}
Electrophysiological source imaging is severely underdetermined due to the lead-field matrix has limited rank relative to the source dimension and strongly correlated columns. Scalp EEG measurements constrain only the observable row-space component, leaving a high-dimensional null space unobservable. We propose a two-stage framework coupling posterior diffusion refinement with deterministic null-space completion. A minimum-norm-initialized, singular-coordinate Posterior-DDIM sampler combines a spatiotemporal diffusion prior with direction-dependent measurement reliability to refine the observable component. A frozen ridge predictor and nonlinear temporal residual network then estimate the null component under explicit projection constraints, preserving the refined row estimate and its predicted EEG exactly. All learned components are trained exclusively on simulated cortical activity and its row–null decomposition, without paired EEG–source training data. Across simulations with varying noise levels, the method achieved higher frame-level area under the receiver operating characteristic curve and average precision, together with lower localization error, than conventional and learning-based baselines. Exploratory evaluations on evoked EEG and simultaneous scalp EEG–stereo-EEG recordings showed condition- and event-dependent concordance with cross-modal and intracranial references.
\end{abstract}


\section{Introduction}
\label{sec:introduction}
\IEEEPARstart{E}{lectroencephalographic} source imaging (ESI) aims to reconstruct cortical activity from scalp electroencephalography (EEG), combining the millisecond temporal resolution of EEG with spatially resolved representations of brain dynamics~\cite{Michel2004,He2018,Michel2019}. It has been widely used for epileptogenic source characterization, network analysis, and noninvasive functional mapping, although its accuracy remains strongly influenced by the forward model and the assumptions imposed on the source distribution \cite{Kaiboriboon2012,SohrabpourHe2021}. The fundamental difficulty of ESI lies in its severe underdetermination. The number of candidate cortical sources substantially exceeds the number of EEG sensors, such that many distinct source configurations can produce identical or nearly identical measurements \cite{Grech2008}. Classical ESI methods address this ambiguity through different structural assumptions. Minimum-norm estimators and their standardized variants favor distributed, low-energy solutions \cite{Hamalainen1994,PascualMarqui2002,Dale2000}, while beamforming methods localize activity through adaptive spatial filtering \cite{VanVeen1997}. Sparse and mixed-norm formulations promote more focal source configurations \cite{Uutela1999,Gramfort2012,Ding2009}, whereas extent-oriented approaches introduce spatial structure to better recover distributed activation \cite{Sohrabpour2016,yang2024rejuvenating}. While these assumptions make the inverse problem tractable, they necessarily introduce information that is not directly determined by the measurements \cite{Hauk2004}. Recovering physiologically meaningful source activity therefore requires assumptions about source structure in addition to the EEG measurements.

Data-driven approaches have emerged as a flexible alternative for capturing source structures that are difficult to encode through hand-crafted regularization. Early deep learning approaches explored spatiotemporal recurrent modeling and convolutional source estimation \cite{Dinh2021,Hecker2021}, while subsequent methods incorporated biophysical dynamics and physics-informed reconstruction \cite{Sun2022,Morik2024}. More recent studies have increasingly embedded the structure of the inverse problem into learnable optimization frameworks \cite{Jiao2024,Liu2025ADMMESI,Wang2026VSSI,Liu2026DADMM}. Most of these approaches, however, rely on a single learned mapping to perform measurement inversion and source regularization simultaneously. As a result, the reconstructed activity does not explicitly distinguish information supported by the EEG from structure introduced by the learned prior. Diffusion models provide a natural alternative by separating prior learning from measurement conditioning. An unconditional diffusion model can learn spatiotemporal source distributions independently of EEG observations \cite{Ho2020,Song2021}, while the known forward operator can be incorporated only at inference time without retraining the generative prior \cite{Chung2023,Kawar2022,Jiao2026,Wang2023}. The remaining challenge is how to exploit such a prior when the lead field constrains only a low-dimensional, strongly anisotropic observable subspace, while a large portion of cortical activity lies in its null space.

This distinction is central to how a diffusion prior should be used for ESI. Since the EEG likelihood depends only on the projection of cortical activity onto the row space of the lead field, measurement guidance can directly constrain only the observable component, whereas any null-space perturbation leaves the predicted EEG unchanged. Most diffusion-based inverse solvers do not explicitly separate these two roles. In diffusion posterior sampling (DPS) \cite{Chung2023}, for example, likelihood guidance can affect the null component only indirectly through the Jacobian of the learned denoiser, leaving measurement-equivalent null-space completions primarily determined by the learned prior. Measurement-consistent methods such as denoising diffusion restoration models (DDRM) and Posterior-DDIM \cite{Kawar2022,Jiao2026} provide more explicit conditioning along observable singular directions, but data consistency alone does not ensure correct cortical localization. Distinct source configurations can remain equally consistent with the EEG while differing substantially in spatial support. Moreover, the observable directions themselves are not equally informative. The singular values of a realistic lead field span more than two orders of magnitude (Section~\ref{sec:formulation}), making weakly observed directions considerably more sensitive to measurement noise.

These limitations motivate treating the row and null components as distinct estimation problems rather than resolving both through a single diffusion trajectory. Range--null formulations make this separation explicit. The denoising diffusion null-space model (DDNM) assigns the measurement-determined range component through the pseudoinverse and uses the generative prior to recover the null component \cite{Wang2023}. For an ill-conditioned lead field, however, direct pseudoinverse recovery can strongly amplify noise along weak singular directions. A posterior factorization into a row-space term and a conditional null-space term \cite{Guo2026} further clarifies how the unobservable component should be estimated for localization. Once the row component is given, the EEG provides no additional information about the null component, and the minimum mean-square-error point estimate is therefore its conditional mean. In contrast, an individual posterior sample retains conditional variability that can introduce unsupported spatial activity into a localization map (Section~\ref{sec:formulation}). These observations suggest assigning the two subspaces to different estimators, with reliability-aware inference in the observable row space and deterministic conditional completion in the null space.

\begin{figure*}[t]
\centering
\includegraphics[width=0.8\textwidth]{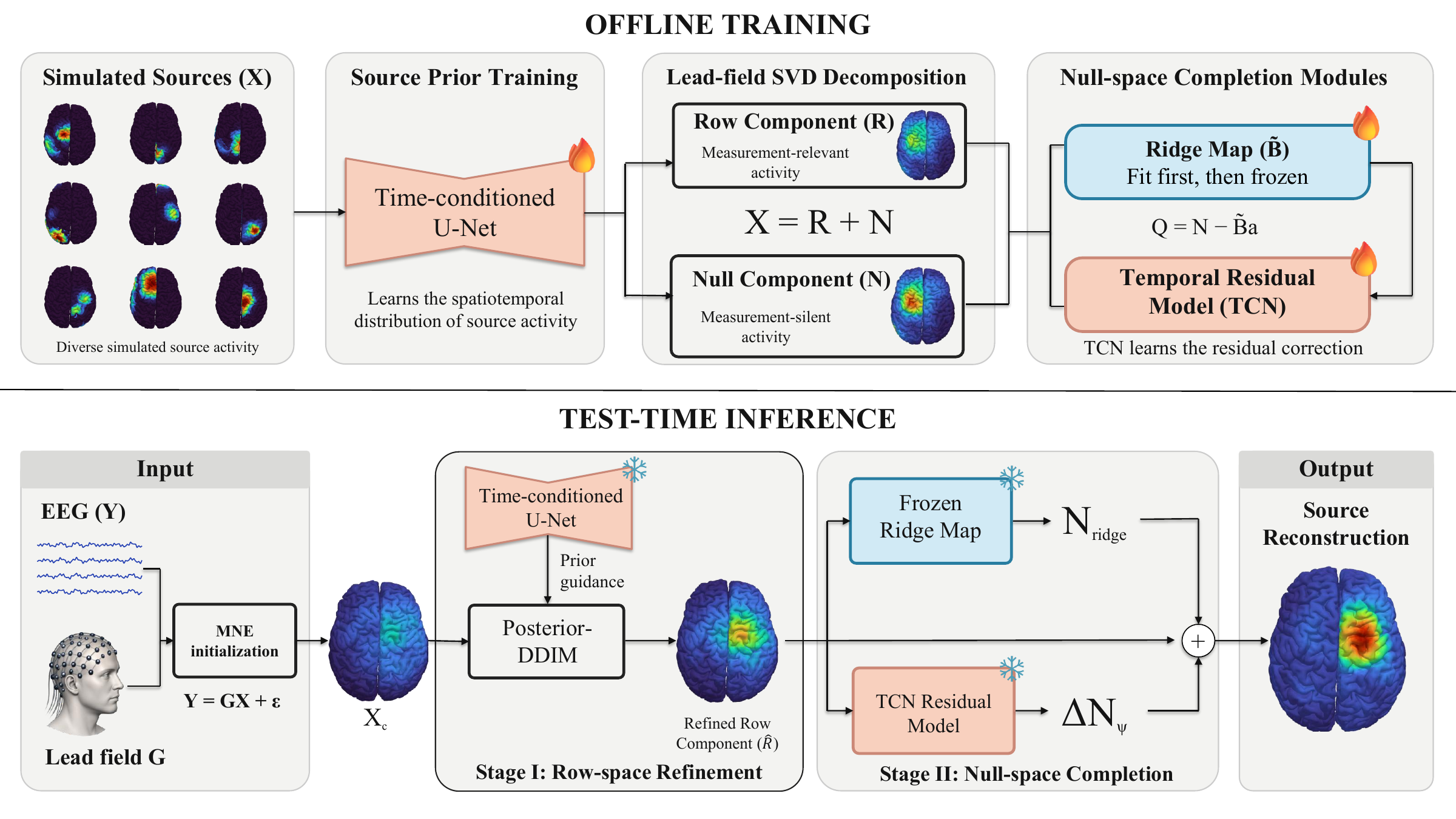}
\caption{Overview of the observability-aligned framework. Offline training (top): a spatiotemporal diffusion prior is trained on clean simulated sources. The singular value decomposition (SVD) of the lead field splits each source into row and null components, from which a ridge map and a residual TCN acting in a null-space basis are fitted. Test-time inference (bottom): Stage I refines the MNE-initialized row component with singular-coordinate Posterior-DDIM. Stage II adds the frozen ridge and TCN null-space completions, which leave the predicted EEG unchanged.}
\label{fig:framework}
\end{figure*}

Accordingly, we formulate learned-prior ESI as two observability-aligned estimation tasks (Fig.~\ref{fig:framework}). Stage I reconstructs the row-space component using a singular-coordinate Posterior-DDIM sampler initialized from the minimum-norm estimate (MNE), balancing measurement reliability along each observable direction against the learned source prior. Stage II then performs deterministic null-space completion following the conditional-mean principle above. A frozen ridge map captures the linearly predictable component, while a temporal convolutional network (TCN), trained on clean row--null pairs, predicts the remaining nonlinear dependence. Both completion terms are explicitly projected onto the lead-field null space and therefore cannot modify either the refined row estimate or its predicted EEG. The resulting reconstruction combines measurement-conditioned row-space refinement with deterministic, prior-driven null-space completion to produce a source estimate for localization.

The main contributions of this work are summarized as follows:
\begin{enumerate}
\item \emph{Source-only prior learning.} All learned components, including the diffusion prior, ridge map, residual basis, and TCN, are trained exclusively on simulated cortical activity and its lead-field-defined row--null decomposition, without using EEG measurements or paired EEG--source training samples. This design separates source-prior learning from measurement-conditioned inversion.
\item \emph{Observability-aligned diffusion reconstruction.} An MNE-initialized, singular-coordinate Posterior-DDIM sampler combines the learned source prior with direction-dependent measurement reliability to refine the observable row-space component. The reverse process operates in the full source space, while only the refined row component is retained for subsequent reconstruction.
\item \emph{Deterministic, measurement-preserving null-space completion.} Motivated by conditional-mean estimation, a frozen ridge predictor and a TCN-based nonlinear residual complete the unobservable component from the refined row estimate. Both terms are confined to the lead-field null space, ensuring that neither the refined row component nor its predicted EEG is altered.
\end{enumerate}


\section{Related Work}
\label{sec:related}
\subsection{Conventional Priors for EEG Source Imaging}
Under the linear forward model, scalp EEG is a lead-field projection of cortical activity corrupted by noise. Since many cortical configurations can explain the same measurements, conventional ESI methods resolve this ambiguity through explicit structural assumptions, many of which admit a unified Bayesian interpretation through different source covariance models and inference rules \cite{Wipf2009}. Minimum-norm estimation (MNE) favors distributed low-energy solutions \cite{Hamalainen1994}, while sLORETA and dSPM standardize or noise-normalize the estimate \cite{PascualMarqui2002,Dale2000}, and LCMV performs localization through adaptive spatial filtering \cite{VanVeen1997}. Sparse and mixed-norm formulations promote focal or temporally structured activity \cite{Uutela1999,Gramfort2012}, whereas extent-oriented methods impose structure on spatial variations, source boundaries, or cortical geometry to better recover distributed generators \cite{Ding2009,Sohrabpour2016,yang2024rejuvenating}. These approaches are interpretable and require no learned training distribution, but their priors are analytically specified and offer limited flexibility for modeling complex spatiotemporal source dynamics. In our framework, MNE serves as an observation-informed initialization for row-space refinement and as the basis of the residual noise-level heuristic in Section~\ref{sec:configuration}.

\subsection{Deep Learning for EEG Source Imaging}
Deep learning methods replace hand-crafted priors with source structure learned from simulated data. Representative approaches include convolutional source estimation \cite{Hecker2021}, spatiotemporal recurrent modeling \cite{Dinh2021}, neural-mass-informed networks \cite{Sun2022}, and physics-informed volumetric reconstruction \cite{Morik2024}. A complementary line of work embeds iterative solvers into trainable architectures. XDL-ESI learns spatiotemporal regularization through unfolded optimization \cite{Jiao2024}, while ADMM-ESINet, VSSI2p-Net, and D-ADMM-Net incorporate learned sparsity, variation, or diffusion-based denoising into unfolded ADMM updates \cite{Liu2025ADMMESI,Wang2026VSSI,Liu2026DADMM}.

Despite their architectural differences, these methods do not explicitly separate source reconstruction according to lead-field observability. Measurement-supported structure and information introduced by the learned model therefore remain coupled. Our framework instead separates these roles at the subspace level. An unconditional diffusion prior is coupled with the forward model only during row-space refinement, while the null component is completed separately and confined to the lead-field null space. This preserves the refined row estimate and its predicted EEG, while adapting measurement conditioning across singular directions according to their reliability.

\subsection{Diffusion Priors for Linear Inverse Problems}
Denoising diffusion probabilistic models (DDPMs) learn a generative prior by reversing progressive noising \cite{Ho2020}, while denoising diffusion implicit models (DDIMs) enable accelerated deterministic or partially stochastic sampling \cite{Song2021}. For linear inverse problems, DPS incorporates likelihood gradients into the reverse process \cite{Chung2023}, DDRM exploits the singular structure of the forward operator \cite{Kawar2022}, and DDNM separates range-space consistency from null-space generation \cite{Wang2023}. Posterior-DDIM further derives singular-coordinate posterior transitions that balance measurement and prior information according to direction-wise reliability \cite{Jiao2026}. These methods provide principled measurement conditioning, but they do not by themselves resolve ambiguity among measurement-equivalent cortical supports. We therefore use Posterior-DDIM only to refine the observable row-space component, with MNE initialization and direction-dependent reliability, while assigning the null component to a separate deterministic completion stage (Section~\ref{sec:method}).

\subsection{Observable--Null Decomposition}
The decomposition of inverse solutions into observable and null-space components is well established in inverse-problem theory. In classical M/EEG analysis, the minimum-norm pseudoinverse recovers the component determined by ideal noiseless measurements, whereas the null-space component requires additional assumptions \cite{Hauk2004}. Learned inverse methods have incorporated the same geometry through null-space networks that preserve predicted measurements \cite{Schwab2019}, explicit range--null decomposition \cite{Chen2020Decomposition}, and diffusion restoration with range-consistent and prior-driven null components \cite{Wang2023}. More recently, Guo \emph{et al.} factorized the inverse posterior into an identifiable component and a conditional null-space distribution to characterize intrinsic ambiguity \cite{Guo2026}.\\

Building on this decomposition, our framework assigns the two subspaces different estimators according to their observability. The noisy and anisotropically observed row component is refined through singular-coordinate posterior diffusion, whereas the unobservable null component is completed deterministically using a conditional-mean-motivated estimator. This separation exploits measurement reliability where EEG is informative while producing a deterministic null-space completion for localization.

\section{Method}
\label{sec:method}
\subsection{Problem Formulation and Observability Decomposition}
\label{sec:formulation}
Let $X=[x_1,\ldots,x_T]\in\mathbb{R}^{N_s\times T}$ denote the cortical activity at $N_s$ source nodes over $T$ frames, and let $Y\in\mathbb{R}^{N_c\times T}$ denote the EEG recorded by $N_c$ channels:
\begin{equation}
Y=GX+E,\qquad E\sim\mathcal{N}(0,\sigma_y^2I).
\label{eq:forward}
\end{equation}
Let $G=U\Sigma V^\top$ be the compact SVD of the lead field $G$, where $r$ is its rank, $\Sigma=\operatorname{diag}(s_1,\ldots,s_r)$ with $s_1\ge\cdots\ge s_r>0$, and $u_j$ and $v_j$ are the $j$th columns of $U$ and $V$. We define
\begin{equation}
P_R=VV^\top,\quad P_N=I-P_R,\quad R=P_RX,\quad N=P_NX.
\label{eq:projection}
\end{equation}
Then $X=R+N$ and $GP_N=0$, so that $p(Y\mid X)=p(Y\mid R)$. In other words, $Y$ is conditionally independent of $N$ given $R$, and hence
\begin{equation}
p(R,N\mid Y)=p(R\mid Y)\,p(N\mid R).
\label{eq:posteriorfactor}
\end{equation}
This factorization organizes the method. Stage I estimates the row component from $p(R\mid Y)$, and Stage II completes the null component from $p(N\mid R)$, which does not involve the EEG. The final source waveform is their sum.

The remaining question for Stage II is which completion to report. For any completion $f(R)$ with finite second moments,
\begin{equation}
\begin{aligned}
\mathbb{E}\|N-f(R)\|_F^2
&=\mathbb{E}\operatorname{tr}\operatorname{Cov}(N\mid R)\\
&\quad+\mathbb{E}\bigl\|\mathbb{E}[N\mid R]-f(R)\bigr\|_F^2,
\end{aligned}
\label{eq:mmse}
\end{equation}
so the conditional mean $\mathbb{E}[N\mid R]$ minimizes the expected squared completion error. By contrast, a draw $\widetilde N\sim p(N\mid R)$ that is conditionally independent of $N$ attains $2\,\mathbb{E}\operatorname{tr}\operatorname{Cov}(N\mid R)$, twice the minimum. Stage II therefore approximates the conditional mean deterministically, first with a linear map (Section~\ref{sec:ridge}) and then with a learned nonlinear residual (Section~\ref{sec:tcn}). Squared error is not a localization metric, and Stage II receives an estimated rather than exact $R$. Equation~\eqref{eq:mmse} therefore motivates the design, and Section~\ref{sec:solvercomp} tests it empirically.

\begin{figure}[t]
\centering
\includegraphics[width=\columnwidth]{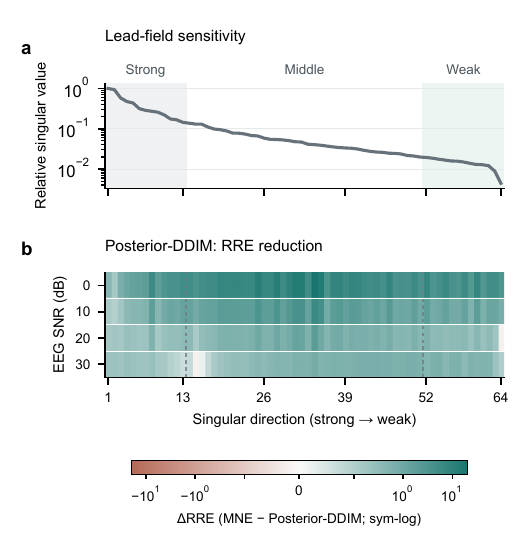}
\caption{Lead-field sensitivity and directional reconstruction error. (a) Relative singular spectrum. (b) Mean paired RRE reduction (MNE minus the updated Posterior-DDIM implementation) on 60 validation identities, four SNRs, and four noise repeats (960 observations), before null-space completion. Repeats are averaged within identity. Teal indicates lower error; the color scale is symmetric logarithmic. Strong, middle, and weak groups comprise directions 1--13, 14--51, and 52--64.}
\label{fig:spectrum}
\end{figure}

Multiplying \eqref{eq:forward} by $\Sigma^{-1}U^\top$ yields the sufficient observed coordinates
\begin{equation}
\begin{aligned}
\bar Y&=\Sigma^{-1}U^\top Y=V^\top X+\widetilde E,\\
\operatorname{Cov}(\widetilde E)&=\sigma_y^2\Sigma^{-2}.
\end{aligned}
\label{eq:sufficient}
\end{equation}
Each singular direction thus has its own effective noise level, $\sigma_y/s_j$ (Fig.~\ref{fig:spectrum}a). This coordinate-dependent uncertainty motivates the posterior transition described below.

\subsection{Learning a Spatiotemporal Source Prior}
Each clean training sequence is divided by its absolute peak to give $X_0$. Under a cosine noise schedule, the forward diffusion process is
\begin{equation}
X_t=\alpha_tX_0+\sigma_t\epsilon,\qquad \epsilon\sim\mathcal{N}(0,I),
\label{eq:diffusion}
\end{equation}
with $\alpha_t^2+\sigma_t^2=1$. The denoising network operates jointly over time and cortical nodes and is trained with $v$-prediction; its architecture and training configuration are given in Section~\ref{sec:configuration}. From the network output $v_\theta(X_t,t)$, the clean-source and noise predictions are
\begin{equation}
\begin{aligned}
\widehat X_0&=\alpha_tX_t-\sigma_tv_\theta(X_t,t),\\
\widehat\epsilon&=\sigma_tX_t+\alpha_tv_\theta(X_t,t).
\end{aligned}
\label{eq:vprediction}
\end{equation}
Training minimizes $\mathcal{L}_v+\lambda_{\mathrm{top}}\mathcal{L}_{\mathrm{top}}$, where $\mathcal{L}_v$ is the standard $v$-prediction loss. Let $\Delta_X\in\mathbb{R}^{N_bT\times N_s}$ stack the clean-source reconstruction errors over a minibatch of $N_b$ sequences and their $T$ frames. With the fixed $N_s\times N_s$ source-relation (topology) matrix $K$ of XDL-ESI \cite{Jiao2024}, the regularizer is
\begin{equation}
\mathcal{L}_{\mathrm{top}}=\frac{1}{N_s^2}
\left\|\left(\frac{\Delta_X^\top \Delta_X}{N_bT}\right)\odot K\right\|_F^2.
\label{eq:topology}
\end{equation}
It penalizes the relation-weighted covariance of reconstruction errors and thus discourages structured errors between related cortical nodes.

\subsection{Singular-Coordinate Posterior Refinement of the Row Component}
Stage I estimates the row-space component using the posterior factor $p(R\mid Y)$ in \eqref{eq:posteriorfactor}. Let $a=V^\top R=V^\top X\in\mathbb{R}^{r\times T}$ denote the observable coordinates, so that $R=Va$. Let $a_j$ and $\bar y_j=u_j^\top Y/s_j$ denote the $j$th rows of $a$ and $\bar Y$. In these coordinates, \eqref{eq:sufficient} becomes
\begin{equation}
\bar y_j=a_j+\widetilde e_j,\qquad
\widetilde e_j\sim\mathcal{N}\Bigl(0,\frac{\sigma_y^2}{s_j^2}I_T\Bigr).
\label{eq:coordinateobservation}
\end{equation}
The direction-specific variance $\sigma_y^2/s_j^2$ makes observability anisotropic even within the row space. Directions with large $s_j$ are strongly constrained by the EEG, whereas directions with small $s_j$ must rely more on the learned source prior. Stage I therefore balances measurement and prior information within the row space rather than assigning the entire row space to the measurement.

We adopt Posterior-DDIM \cite{Jiao2026} because its singular-coordinate formulation matches this lead-field geometry. At reverse step $t$, it compares the measurement signal-to-noise ratio $s_j/\sigma_y$ with the diffusion signal-to-noise ratio $\alpha_t/\sigma_t$. With the log signal-to-noise ratio $\lambda_t=\log(\alpha_t/\sigma_t)$, the observable coordinates are partitioned as
\begin{equation}
\begin{aligned}
\mathcal{M}_t&=\{j:\alpha_t\sigma_y<\sigma_t s_j\},\\
\mathcal{P}_t&=\{1,\ldots,r\}\setminus\mathcal{M}_t.
\end{aligned}
\label{eq:snrpartition}
\end{equation}
The partition follows from variance matching. The scaled observation $\alpha_t\bar y_j$ already carries noise variance $\alpha_t^2\sigma_y^2/s_j^2$. It can therefore serve as a diffusion state at time $t$ only when the residual variance $\sigma_t^2-\alpha_t^2\sigma_y^2/s_j^2$ is positive, which is exactly the condition $j\in\mathcal{M}_t$; equality defines the crossing time $\kappa_j$ introduced below. Coordinates $j\in\mathcal{P}_t$ remain observable, but they cannot be anchored directly to the measurement, whose noise exceeds the diffusion noise at that time. The two sets are time-dependent update regimes rather than fixed classes of directions. The crossing time $\kappa_j$, defined by $\lambda_{\kappa_j}=\log(s_j/\sigma_y)$, is the time at which measurement and diffusion reliabilities are equal. At this time, $x_{\kappa_j,j}=\alpha_{\kappa_j}\bar y_j$, which connects the two regimes.

Let $\mathcal{D}_{\eta,t\to t'}(x,\mu,z)$ denote the DDIM transition \cite{Song2021} from time $t$ to $t'$ with stochasticity $\eta$, clean prediction $\mu$, and innovation $z$. Because $G$ acts on each frame independently, $x_{t,j}=v_j^\top X_t$ and $\mu_{t,j}=v_j^\top\widehat X_0$ are row vectors over time. Following Posterior-DDIM \cite{Jiao2026}, the prior-dominated branch keeps this source-prior transition but injects the measurement through the crossing-conditioned clean prediction
\begin{equation}
\mu^{\mathrm{corr}}_{t,j}=\mu_{t,j}+
\frac{\bar y_j-x_{t,j}/\alpha_t}
{\exp\{2(\lambda_t-\lambda_{\kappa_j})\}-1}.
\label{eq:pddimcorrection}
\end{equation}
We draw $\zeta_j\sim\mathcal{N}(0,I_T)$ once for the measurement trajectory and denote the coordinate-wise and full-space DDIM innovations by $z_{t',j}$ and $Z_{t'}$, respectively. One reverse step $t\to t'$ then has three regimes:
\begin{equation}
\begin{aligned}
&j\in\mathcal{M}_{t'}\ \text{(measurement-dominated)}:\\
&\qquad x_{t',j}=\alpha_{t'}\bar y_j
+\sqrt{\sigma_{t'}^2-\alpha_{t'}^2\sigma_y^2/s_j^2}\;\zeta_j,\\[1mm]
&j\in\mathcal{P}_{t'}\ \text{(prior-dominated)}:\\
&\qquad x_{t',j}=\mathcal{D}_{\eta_{\mathrm{obs}},t\to t'}
\bigl(x_{t,j},\mu^{\mathrm{corr}}_{t,j},z_{t',j}\bigr),\\[1mm]
&\text{null space (prior only)}:\\
&\qquad P_NX_{t'}=\mathcal{D}_{\eta_{\mathrm{null}},t\to t'}
\bigl(P_NX_t,P_N\widehat X_0,P_NZ_{t'}\bigr).
\end{aligned}
\label{eq:regimeupdates}
\end{equation}
The regime is selected at the destination time $t'$. In the discrete solver, a step that crosses $\kappa_j$ is taken directly, without an intermediate crossing state or an additional denoiser evaluation. Equation~\eqref{eq:pddimcorrection} is evaluated at the current grid point and the correction is set to zero when its denominator falls below $10^{-7}$ in magnitude. These rules define the finite-step approximation used for the direction-dependent posterior updates.

Because ESI is severely underdetermined, we warm-start the reverse process from the MNE solution $X_c$, which provides a stable, measurement-compatible configuration. With $c=\|X_c\|_\infty$, we divide $X_c$, $Y$, and $\sigma_y$ by $c$ and initialize
\begin{equation}
X_\tau=\alpha_\tau X_c/c+\sigma_\tau Z,\qquad Z\sim\mathcal{N}(0,I).
\label{eq:warm}
\end{equation}
This warm start replaces the terminal Gaussian initialization, so the exact-sampler consistency result of \cite{Jiao2026} does not carry over unchanged.

The reverse process evolves the full source state, allowing the denoiser to use joint row–null structure when updating the observable component. Let $X_{\mathrm{end}}$ denote the final state of the reverse process. We restore the source scale using (c) from (13) and project onto the lead-field row space to obtain
\begin{equation}
\widehat R=c\,P_RX_{\mathrm{end}}.
\label{eq:refinedrow}
\end{equation}
The resulting row estimate $\hat R$ is then passed to Stage II for deterministic null-space completion.

\subsection{Ridge-Based Null-Space Completion}
\label{sec:ridge}
Stage II first exploits population-level row--null dependence to obtain a stable linear approximation of $\mathbb{E}[N\mid R]$ that does not change the predicted EEG. This map captures the linearly predictable part of the null component and serves as the reference for subsequent residual learning.

For training sequence $k$, let $a_k=V^\top X_k$ and $N_k=P_NX_k$. Ridge regression \cite{Hoerl1970} on clean training sources gives
\begin{equation}
B_\gamma=\operatorname*{arg\,min}_B\,\sum_k\|N_k-Ba_k\|_F^2+\gamma\|B\|_F^2.
\label{eq:ridge}
\end{equation}
The fitted map is projected onto the null space, $\widetilde B=P_NB_\gamma$, and applied to the estimated coordinates $\widehat a=V^\top\widehat R$:
\begin{equation}
\widehat N_{\mathrm{ridge}}=\widetilde B\widehat a,\qquad
\widehat X_{\mathrm{ridge}}=\widehat R+\widehat N_{\mathrm{ridge}}.
\label{eq:context}
\end{equation}
Because $G\widehat N_{\mathrm{ridge}}=0$, the ridge-completed waveform satisfies $G\widehat X_{\mathrm{ridge}}=G\widehat R$. The map remains frozen during subsequent residual learning. Because $\widetilde B$ is linear and acts on each frame $a_t$ independently, it cannot represent nonlinear row--null dependence. This limitation motivates a learned nonlinear residual correction.

\subsection{Nonlinear Residual Completion in the Null Space}
\label{sec:tcn}
The remaining task is to predict the null-space structure that the linear, framewise ridge map leaves unexplained. For clean training sequence $k$, the residual target is
\begin{equation}
Q_k=P_NX_k-\widetilde Ba_k.
\label{eq:residual_target}
\end{equation}
Rather than predicting an unrestricted cortical residual, we restrict the correction to a low-dimensional spatial basis derived from these training residuals. We concatenate the residuals across training sequences and frames and retain their $d$ leading uncentered left singular vectors. Projection by $P_N$ and orthonormalization then give a fixed basis $W\in\mathbb{R}^{N_s\times d}$ with $V^\top W=0$. This basis spans the correction directions supported by the training data, and the time-varying coefficients of these directions are predicted from the row trajectory.

A residual network $h_\psi$ maps the sequence of row coordinates to a sequence of residual coefficients. We implement it as a TCN based on dilated residual temporal convolutions \cite{Bai2018TCN}, but replace causal padding with symmetric padding. Its noncausal convolutions use both preceding and subsequent frames within the receptive field, so the model performs offline reconstruction rather than causal forecasting. A framewise network without temporal context is compared in Section~\ref{sec:ablation}. Let $s_X$ denote the global RMS amplitude of the training sources, and let $\Lambda_a$ be the stabilized diagonal matrix of coordinate-wise RMS values of $a_k/s_X$. The predicted coefficients and the correction in source units are
\begin{equation}
\begin{aligned}
C_\psi(a)&=h_\psi(\Lambda_a^{-1}a/s_X)\in\mathbb{R}^{d\times T},\\
\Delta N_\psi(a)&=s_XP_NW C_\psi(a).
\end{aligned}
\label{eq:tcn_residual}
\end{equation}
Because $V^\top W=0$, the projector $P_N$ acts as the identity on $W$ and is retained only to suppress round-off error. The basis restricts which residual patterns can be expressed, whereas the network predicts their framewise contributions from the row trajectory within its receptive field. The residual supplements the ridge prediction; it neither replaces that prediction nor modifies the row estimate. Section~\ref{sec:configuration} summarizes the architecture.

Only the TCN is optimized, using a source-space Huber loss \cite{Huber1964} on clean training pairs:
\begin{equation}
\mathcal{L}_{\mathrm{res}}=\frac{1}{MN_sT}\sum_{k,i,t}
\rho_\delta\!\left(\frac{[\Delta N_\psi(a_k)-Q_k]_{it}}{s_X}\right),
\label{eq:residual_loss}
\end{equation}
where $M$ is the number of training sequences and $\rho_\delta$ is the Huber penalty with transition parameter $\delta$. The Huber loss reduces the influence of large residual errors and targets a robust conditional location, which can differ from the conditional mean. The diffusion prior, ridge map, residual basis, and normalization statistics remain fixed during this optimization. Clean row coordinates are used only to learn the conditional completion; at inference, they are replaced by the estimate $\widehat a=V^\top\widehat R$ obtained from the EEG.

The final output is the time-resolved source waveform
\begin{equation}
\widehat X=\widehat R+\widetilde B\widehat a+
\Delta N_\psi(\widehat a).
\label{eq:final_reconstruction}
\end{equation}
Because both completion terms lie in the null space,
\begin{equation}
P_R\widehat X=\widehat R,\qquad G\widehat X=G\widehat R.
\label{eq:completion_preservation}
\end{equation}
The completion therefore preserves the upstream row estimate and its predicted EEG, following the measurement-preserving principle of null-space learning \cite{Schwab2019}.

\section{Experiments and Results}
\label{sec:experiments}
\label{sec:results}
\subsection{Data, Metrics, and Implementation}
\label{sec:setup}
\label{sec:configuration}
We simulated 128-frame source sequences with The Virtual Brain (TVB) \cite{SanzLeon2013} on $N_s=2{,}052$ cortical nodes, each containing one to six stimulated centers followed by propagated dynamics. EEG was generated with a fixed $64\times2{,}052$ lead field and white Gaussian noise at SNRs of 0, 10, 20, and 30~dB, defined on Frobenius norms. We call each source sequence an identity and each EEG realization at one SNR an observation. The 12,000 identities were split into 8,000 training, 2,000 validation, and 2,000 test identities. The source prior, ridge map, residual basis, and TCN were trained on clean training identities, and the learned baselines were retrained on simulated EEG at all four SNRs. Unless otherwise stated, all methods were evaluated on the 2,000 test identities (8,000 observations). In simulation, Stage I used the generating noise level.

Localization was scored frame by frame on the active frames $\mathcal{T}$, whose ground-truth energy $e_t=\|x_t\|_2^2$ is at least 10\% of the sequence maximum. In frame $t$, the ground-truth and predicted supports are
\begin{equation}
\begin{aligned}
\Omega_t&=\{i:|X_{it}|\ge0.1\max\nolimits_j|X_{jt}|\},\\
\widehat\Omega_t&=\{i:|\widehat X_{it}|\ge0.1\max\nolimits_j|\widehat X_{jt}|\}.
\end{aligned}
\label{eq:support}
\end{equation}
Frame AUC is the area under the receiver operating characteristic curve obtained by ranking all nodes by $|\widehat X_{it}|$ against their membership in $\Omega_t$, and average precision (AP) is the area under the corresponding precision--recall curve, with tied scores treated as a single threshold. Both are therefore independent of the prediction threshold. The F1 score, $2|\Omega_t\cap\widehat\Omega_t|/(|\Omega_t|+|\widehat\Omega_t|)$, measures support overlap. The symmetric localization error (LE) averages the directed mean nearest-neighbor geodesic distances between the two supports,
\begin{equation}
\begin{aligned}
\mathrm{LE}_t&=\tfrac12\bigl[\bar g(\Omega_t,\widehat\Omega_t)+\bar g(\widehat\Omega_t,\Omega_t)\bigr],\\
\bar g(\mathcal{A},\mathcal{B})&=\frac{1}{|\mathcal{A}|}\sum_{i\in\mathcal{A}}\min_{j\in\mathcal{B}}g_{ij},
\end{aligned}
\label{eq:le}
\end{equation}
where $g_{ij}$ is the cortical geodesic distance, and node pairs in disconnected components are assigned the largest finite distance. LE thus penalizes both missed and spurious activity. Each metric is averaged over $\mathcal{T}$ with weights $e_t$ and then over observations. Tables~\ref{tab:mechanistic}--\ref{tab:stress} report mean $\pm$ sample SD across identities (divisor $n-1$); unless stratified by SNR or evaluated with exact rows, the four SNR scores are first averaged within each identity. 
SD describes between-identity variation under fixed models, not variability across training runs. Paired 95\% confidence intervals, shown in brackets, were obtained from 2,000 bootstrap replicates over identities \cite{Efron1993}.

The source prior is a two-dimensional U-Net with 32 base channels, trained with $v$-prediction on a 1,000-step cosine schedule and $\lambda_{\mathrm{top}}=1000$. Posterior refinement started at $\tau=250$ and used 100 denoiser evaluations, with $\eta_{\mathrm{obs}}=0$ and $\eta_{\mathrm{null}}=1$. The residual basis retained $d=256$ vectors. The TCN comprised six noncausal residual blocks (128 channels, dilations 1--32), whose receptive field spans the whole sequence; its output layer was zero-initialized so that training started from the ridge solution. For recorded EEG, the noise scale was set as $\widehat\sigma_y=\max\{\operatorname{RMS}(GX_c-Y),\,0.05\operatorname{RMS}(Y)\}$.

\subsection{Contribution of Observation-Conditioned Refinement}
\label{sec:stage1res}
We assess recovery along singular direction \(j\) using the relative reconstruction error of its coefficient time series \(a_{j,:}\), defined as
\[
\mathrm{RRE}_j=\frac{\|\widehat a_{j,:}-a_{j,:}\|_2}{\|a_{j,:}\|_2}.
\]
On validation identities, the Posterior-DDIM implementation generally reduced directional RRE relative to MNE across the four SNRs (Fig.~\ref{fig:spectrum}b).

To isolate the role of the measurement updates in Stage I, we compared three row estimates followed by the same frozen ridge + TCN completion: the MNE initializer, prior-only refinement, and full Posterior-DDIM refinement. Prior-only refinement used the same MNE-based warm start and reverse schedule as Posterior-DDIM, but omitted the measurement branch in \eqref{eq:regimeupdates} and the correction in \eqref{eq:pddimcorrection}. Its reverse updates were therefore driven solely by the learned source prior. The comparison covered four SNRs with four independent noise realizations per identity 

Posterior-DDIM achieved the highest AP and the smallest LE. Compared with prior-only refinement, it increased AP by 0.0701 [0.0535, 0.0877] and reduced LE by 4.67~mm [3.45, 6.08]. Compared with MNE, it increased AP by 0.1114 [0.0882, 0.1369] and reduced LE by 23.04~mm [20.49, 25.49]. These results show complementary gains from source-prior refinement and measurement conditioning: prior-only refinement improved on MNE, and incorporating EEG measurements during refinement further improved localization.

\begin{table}[t]
\caption{Stage I inputs followed by the same frozen ridge + TCN completion on validation identities (four SNRs and four noise realizations per identity). Brackets give 95\% identity-cluster bootstrap intervals.}
\label{tab:stage1}\centering\footnotesize
\setlength{\tabcolsep}{3pt}
\begin{tabular}{@{}lrr@{}}
\toprule
Stage I input & AP $\uparrow$ & LE (mm) $\downarrow$\\
\midrule
MNE & 0.3794 [0.3242, 0.4316] & 53.16 [50.00, 56.46]\\
Prior only & 0.4206 [0.3602, 0.4798] & 34.78 [30.41, 39.46]\\
Posterior-DDIM & \textbf{0.4907} [0.4324, 0.5470] & \textbf{30.11} [26.25, 34.36]\\
\bottomrule
\end{tabular}
\end{table}

\subsection{Ablations}
\label{sec:ablation}
Table~\ref{tab:mechanistic} crosses two row inputs with two completions, keeping the completion weights and scoring rules fixed. With ridge + TCN completion, Posterior-DDIM rows improved all four metrics relative to MNE rows at every SNR and reduced LE by 21.82~mm [21.35, 22.32], most strongly at 0~dB (39.08~mm). The refined row was thus the better input for completion. Null-space completion supplied most of the ranking gain, raising frame AUC from 0.8213 for the refined row alone to 0.8977 with ridge completion. The learned residual then mainly improved support overlap and spatial agreement: it increased AP by 0.0644 [0.0593, 0.0695] and reduced LE by 14.67~mm [14.05, 15.30], whereas its AUC gain was not resolved. Using exact clean rows reduced LE from 30.90 to 21.00 mm, showing the dependence of completion accuracy on the row input. In a separate control with a 64-dimensional basis, a framewise multilayer perceptron slightly exceeded the TCN in AP (0.5211 versus 0.5099), indicating that the main benefit stems from nonlinear completion; temporal context did not provide an additional advantage in this setting.

\begin{table*}[t]
\caption{Row-input and completion ablations on the test set at prediction threshold 0.1. Deployment rows use 8,000 EEG observations; the exact-row diagnostic supplies the ground-truth row component of the 2,000 test identities. Bold marks the best deployment value. Values are mean $\pm$ SD across 2,000 identities; four SNRs are averaged within identity for deployment rows.}
\label{tab:mechanistic}\centering\small
\setlength{\tabcolsep}{5pt}
\renewcommand{\arraystretch}{1.12}
\begin{tabular*}{\textwidth}{@{\extracolsep{\fill}}llrrrr@{}}
\toprule
Row input & Completion & AUC $\uparrow$ & AP $\uparrow$ & F1 $\uparrow$ & LE (mm) $\downarrow$\\
\midrule
\multirow{2}{*}{MNE} & Ridge & $0.8305\pm 0.0939$ & $0.3114\pm 0.1954$ & $0.0886\pm 0.0538$ & $65.34\pm 10.07$\\
 & Ridge + TCN & $0.8465\pm 0.0895$ & $0.3661\pm 0.2017$ & $0.1595\pm 0.0993$ & $52.72\pm 13.23$\\
\addlinespace[2pt]
\multirow{2}{*}{Posterior-DDIM} & Ridge & $0.8977\pm 0.0738$ & $0.4252\pm 0.2260$ & $0.1628\pm 0.1080$ & $45.57\pm 17.09$\\
 & Ridge + TCN & $\mathbf{0.8998}\pm 0.0748$ & $\mathbf{0.4896}\pm 0.2312$ & $\mathbf{0.2866}\pm 0.1682$ & $\mathbf{30.90}\pm 17.64$\\
\midrule
\multirow{2}{*}{\emph{Exact row}} & Ridge & $0.9311\pm 0.0614$ & $0.5141\pm 0.2374$ & $0.1992\pm 0.1368$ & $41.19\pm 18.99$\\
 & Ridge + TCN & $0.9385\pm 0.0604$ & $0.5916\pm 0.2414$ & $0.3808\pm 0.2089$ & $21.00\pm 17.30$\\
\bottomrule
\end{tabular*}
\end{table*}

\subsection{Sampled Versus Deterministic Completion}
\label{sec:solvercomp}
On a fixed subset of 60 test identities (240 observations), we compared deterministic completion with the sampler's own null realization and with DDNM+ \cite{Wang2023}, which used the same source prior, MNE initialization, and 100 denoiser evaluations (Table~\ref{tab:solver}). Deterministic completion increased AP by 0.222 [0.168, 0.279] over the sampled null component and by 0.272 [0.214, 0.333] over the full DDNM+ pipeline, whereas the LE differences were not resolved. Consistent with \eqref{eq:mmse}, sampling the null component mainly degraded the ranking of true support nodes.

\begin{table*}[t]
\caption{Sampled versus deterministic null-space completion on a fixed subset of 60 test identities (240 observations). Every pipeline uses 100 denoiser evaluations; completion costs are additional. Values are mean $\pm$ SD across 60 identities; four SNRs are averaged within identity.}
\label{tab:solver}\centering\small
\setlength{\tabcolsep}{5pt}
\renewcommand{\arraystretch}{1.12}
\begin{tabular*}{\textwidth}{@{\extracolsep{\fill}}lrrrr@{}}
\toprule
Method & AUC $\uparrow$ & AP $\uparrow$ & F1 $\uparrow$ & LE (mm) $\downarrow$\\
\midrule
DDNM+ & $0.6190\pm 0.0835$ & $0.1166\pm 0.0829$ & $0.1395\pm 0.0994$ & $40.35\pm 22.42$\\
Full Posterior-DDIM & $0.7679\pm 0.0865$ & $0.1671\pm 0.1206$ & $0.1566\pm 0.0947$ & $38.87\pm 16.72$\\
Posterior-DDIM row only & $0.8274\pm 0.0874$ & $0.1726\pm 0.1376$ & $0.1389\pm 0.0987$ & $42.06\pm 19.72$\\
Row + Ridge & $\mathbf{0.8922}\pm 0.0801$ & $0.3550\pm 0.2547$ & $0.1344\pm 0.1054$ & $46.56\pm 18.86$\\
Row + Ridge + TCN & $0.8781\pm 0.1071$ & $\mathbf{0.3887}\pm 0.2671$ & $\mathbf{0.2059}\pm 0.1582$ & $\mathbf{38.61}\pm 21.95$\\
\bottomrule
\end{tabular*}
\end{table*}

\subsection{Comparison with Baselines}
\label{sec:baselines}
We compared four learned baselines, ConvDip \cite{Hecker2021}, XDL-ESI \cite{Jiao2024}, ADMM-ESINet \cite{Liu2025ADMMESI}, and VSSI2p-Net \cite{Wang2026VSSI}, reimplemented and retrained following the original paper, and four conventional baselines, MNE \cite{Hamalainen1994}, sLORETA \cite{PascualMarqui2002}, LCMV \cite{VanVeen1997}, and a framewise ADMM solver with source and graph-variation sparsity (Structured ADMM) \cite{Ding2009,Sohrabpour2016}. As shown in Table~\ref{tab:benchmark}, the proposed method attained the highest AUC and AP and the smallest LE at all four SNRs, and the highest F1 at 0--20~dB. Its overall AP exceeded that of the strongest baseline in AP, ConvDip, by 0.0463 [0.0397, 0.0530], although the margin at 10~dB was not resolved. At 30~dB, Structured ADMM attained a higher F1, whereas the proposed method retained a 5.12~mm smaller LE.

\begin{table*}[t]
\caption{Localization benchmark by SNR. Values are mean $\pm$ SD across 2,000 test identities per SNR (8,000 observations per method). Higher AUC, AP, and F1 and lower symmetric LE (mm) are better. Bold marks the best mean in each column.}
\label{tab:benchmark}\centering\footnotesize
\setlength{\tabcolsep}{5pt}
\renewcommand{\arraystretch}{1.08}
\begin{tabular*}{\textwidth}{@{\extracolsep{\fill}}llrrrr@{}}
\toprule
SNR & Method & AUC $\uparrow$ & AP $\uparrow$ & F1 $\uparrow$ & LE (mm) $\downarrow$\\
\midrule
\multirow{9}{*}{0 dB} & Ours & $\mathbf{0.8742}\pm0.0953$ & $\mathbf{0.4195}\pm0.2385$ & $\mathbf{0.2329}\pm0.1527$ & $\mathbf{36.73}\pm20.42$\\
 & ConvDip & $0.8210\pm0.1025$ & $0.3282\pm0.1955$ & $0.0970\pm0.0609$ & $66.02\pm9.29$\\
 & XDL-ESI & $0.8212\pm0.1064$ & $0.2666\pm0.1941$ & $0.1961\pm0.1377$ & $45.99\pm18.53$\\
 & VSSI2p-Net & $0.7356\pm0.1142$ & $0.3021\pm0.2192$ & $0.0839\pm0.0575$ & $73.12\pm9.02$\\
 & ADMM-ESINet & $0.7223\pm0.1197$ & $0.1436\pm0.1185$ & $0.1282\pm0.1046$ & $51.48\pm22.91$\\
 & sLORETA & $0.6920\pm0.0938$ & $0.1185\pm0.0984$ & $0.0522\pm0.0297$ & $77.24\pm5.30$\\
 & MNE & $0.6604\pm0.1223$ & $0.0852\pm0.0755$ & $0.0651\pm0.0386$ & $74.67\pm7.78$\\
 & LCMV & $0.7027\pm0.0891$ & $0.0875\pm0.0563$ & $0.0544\pm0.0305$ & $75.86\pm5.50$\\
 & Structured ADMM & $0.5515\pm0.1295$ & $0.0420\pm0.0312$ & $0.0451\pm0.0404$ & $96.13\pm12.02$\\
\midrule
\multirow{9}{*}{10 dB} & Ours & $\mathbf{0.8957}\pm0.0821$ & $\mathbf{0.4664}\pm0.2401$ & $\mathbf{0.2600}\pm0.1643$ & $\mathbf{32.39}\pm19.16$\\
 & ConvDip & $0.8893\pm0.0817$ & $0.4623\pm0.2339$ & $0.1910\pm0.1213$ & $44.73\pm15.54$\\
 & XDL-ESI & $0.8497\pm0.1051$ & $0.3117\pm0.2213$ & $0.2282\pm0.1597$ & $39.95\pm20.18$\\
 & VSSI2p-Net & $0.8260\pm0.0941$ & $0.3945\pm0.2377$ & $0.1113\pm0.0826$ & $61.75\pm14.03$\\
 & ADMM-ESINet & $0.7244\pm0.1208$ & $0.1661\pm0.1268$ & $0.1286\pm0.1049$ & $51.19\pm22.93$\\
 & sLORETA & $0.7822\pm0.0839$ & $0.1996\pm0.1285$ & $0.0649\pm0.0360$ & $69.69\pm7.29$\\
 & MNE & $0.7583\pm0.1046$ & $0.1471\pm0.1117$ & $0.0950\pm0.0578$ & $59.58\pm13.36$\\
 & LCMV & $0.7585\pm0.0853$ & $0.1032\pm0.0672$ & $0.0676\pm0.0371$ & $68.59\pm7.79$\\
 & Structured ADMM & $0.6671\pm0.1353$ & $0.0999\pm0.0802$ & $0.1186\pm0.0941$ & $82.84\pm16.49$\\
\midrule
\multirow{9}{*}{20 dB} & Ours & $\mathbf{0.9105}\pm0.0809$ & $\mathbf{0.5194}\pm0.2468$ & $\mathbf{0.3080}\pm0.1888$ & $\mathbf{28.34}\pm19.86$\\
 & ConvDip & $0.9045\pm0.0763$ & $0.4895\pm0.2390$ & $0.2294\pm0.1411$ & $37.86\pm16.75$\\
 & XDL-ESI & $0.8530\pm0.1050$ & $0.3173\pm0.2246$ & $0.2322\pm0.1622$ & $39.25\pm20.41$\\
 & VSSI2p-Net & $0.8563\pm0.0875$ & $0.4126\pm0.2424$ & $0.1284\pm0.0939$ & $54.53\pm16.50$\\
 & ADMM-ESINet & $0.7228\pm0.1196$ & $0.1684\pm0.1260$ & $0.1276\pm0.1040$ & $51.25\pm22.87$\\
 & sLORETA & $0.8075\pm0.0788$ & $0.2167\pm0.1319$ & $0.0754\pm0.0420$ & $63.28\pm9.86$\\
 & MNE & $0.7936\pm0.0952$ & $0.1642\pm0.1155$ & $0.1119\pm0.0698$ & $50.56\pm17.30$\\
 & LCMV & $0.7702\pm0.0840$ & $0.1058\pm0.0696$ & $0.0724\pm0.0402$ & $65.87\pm9.19$\\
 & Structured ADMM & $0.7963\pm0.1111$ & $0.2534\pm0.1618$ & $0.2484\pm0.1412$ & $56.68\pm17.59$\\
\midrule
\multirow{9}{*}{30 dB} & Ours & $\mathbf{0.9190}\pm0.0807$ & $\mathbf{0.5530}\pm0.2490$ & $0.3457\pm0.2047$ & $\mathbf{26.12}\pm20.47$\\
 & ConvDip & $0.9067\pm0.0756$ & $0.4930\pm0.2397$ & $0.2347\pm0.1435$ & $37.02\pm16.89$\\
 & XDL-ESI & $0.8533\pm0.1050$ & $0.3179\pm0.2250$ & $0.2326\pm0.1625$ & $39.18\pm20.42$\\
 & VSSI2p-Net & $0.8640\pm0.0847$ & $0.4144\pm0.2429$ & $0.1340\pm0.0959$ & $51.59\pm16.84$\\
 & ADMM-ESINet & $0.7202\pm0.1200$ & $0.1681\pm0.1257$ & $0.1245\pm0.1030$ & $51.97\pm22.65$\\
 & sLORETA & $0.8119\pm0.0779$ & $0.2186\pm0.1323$ & $0.0772\pm0.0430$ & $62.22\pm10.20$\\
 & MNE & $0.8004\pm0.0937$ & $0.1664\pm0.1157$ & $0.1143\pm0.0711$ & $49.22\pm17.59$\\
 & LCMV & $0.7717\pm0.0839$ & $0.1060\pm0.0699$ & $0.0730\pm0.0406$ & $65.56\pm9.35$\\
 & Structured ADMM & $0.8643\pm0.0955$ & $0.4106\pm0.1997$ & $\mathbf{0.3657}\pm0.1762$ & $31.24\pm14.29$\\
\bottomrule
\end{tabular*}

\end{table*}

\subsection{Robustness to Noise and Forward-Model Mismatch}
\label{sec:stress}
To test sensitivity to the idealized simulation assumptions, we reran the frozen pipeline without retraining on validation identities at four SNRs under three departures (Table~\ref{tab:stress}): the residual-based noise estimate used for recorded EEG in place of the generating noise level, a fixed 5\% RMS channel-gain error with inference based on the nominal lead field, and correlated sensor noise with a 50\% common-mode component and unchanged marginal variance. Relative to the nominal condition, the estimated noise level reduced AP by 0.0324 [0.0175, 0.0467] and increased LE by 1.73~mm [0.16, 3.15], and the gain error increased LE by 2.58~mm [0.80, 4.26], whereas correlated noise produced no detectable change.

\begin{table*}[t]
\caption{Frozen proposed pipeline under departures from the simulation assumptions, evaluated on validation identities at four SNRs with one noise realization per identity and SNR. No model was retrained. Values are mean $\pm$ SD across 60 identities; four SNRs are averaged within identity.}
\label{tab:stress}\centering\small
\setlength{\tabcolsep}{5pt}
\renewcommand{\arraystretch}{1.12}
\begin{tabular*}{\textwidth}{@{\extracolsep{\fill}}lrrrr@{}}
\toprule
Condition & AUC $\uparrow$ & AP $\uparrow$ & F1 $\uparrow$ & LE (mm) $\downarrow$\\
\midrule
Nominal & $0.9006\pm 0.0637$ & $0.4937\pm 0.2274$ & $0.2888\pm 0.1610$ & $31.06\pm 16.60$\\
Estimated $\sigma_y$ & $0.8972\pm 0.0679$ & $0.4614\pm 0.2448$ & $0.2628\pm 0.1577$ & $32.78\pm 17.85$\\
5\% channel-gain error & $0.8990\pm 0.0654$ & $0.4888\pm 0.2248$ & $0.2678\pm 0.1536$ & $33.63\pm 17.01$\\
Correlated noise & $0.9017\pm 0.0625$ & $0.4931\pm 0.2316$ & $0.2940\pm 0.1656$ & $30.08\pm 17.12$\\
\bottomrule
\end{tabular*}
\end{table*}

\subsection{Qualitative Multisource Comparison}
\begin{figure*}[t]
\centering
\includegraphics[width=0.98\textwidth]{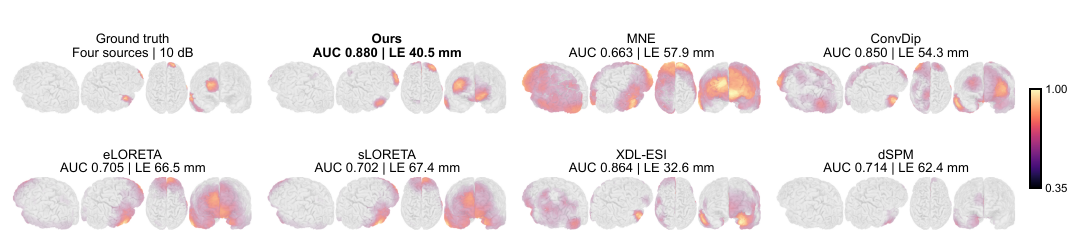}
\caption{Multiview comparison for a four-source simulation at 10~dB. The top row shows the ground truth, Ours, MNE, and ConvDip, and the bottom row shows eLORETA, sLORETA, XDL-ESI, and dSPM. Each map shows the temporal RMS over 128 frames in left-lateral, right-lateral, superior, and posterior views. Interpolated maps are peak-normalized separately within each hemisphere, share one color scale across views, and are colored only above 35\%. AUC and symmetric geodesic LE are computed from the original node waveforms at prediction threshold 0.1, not at the display threshold.}
\label{fig:cases}
\end{figure*}

Fig.~\ref{fig:cases} shows a four-source example at 10~dB. The proposed reconstruction achieved an AUC of 0.880 and an LE of 40.5~mm, compared with 0.663 and 57.9~mm for MNE and 0.850 and 54.3~mm for ConvDip, whereas XDL-ESI achieved a smaller LE of 32.6~mm in this example.

\subsection{MNE Sample Evaluation}
\label{sec:vep_protocol}
\label{sec:vep_results}
We applied the method to the left- and right-ear auditory and left- and right-visual-field responses of the MNE sample participant \cite{MNESampleDataset}, using the canonical fsaverage geometry and a geometry-matched residual basis and TCN trained only on simulated sources. For each method and condition, we took the global source maximum over 29.2--156.2~ms, selected without reference to MEG, and measured its three-dimensional distance to the MEG-dSPM global peak at the same time.

\begin{figure*}[t]
\centering
\includegraphics[width=0.9\textwidth]{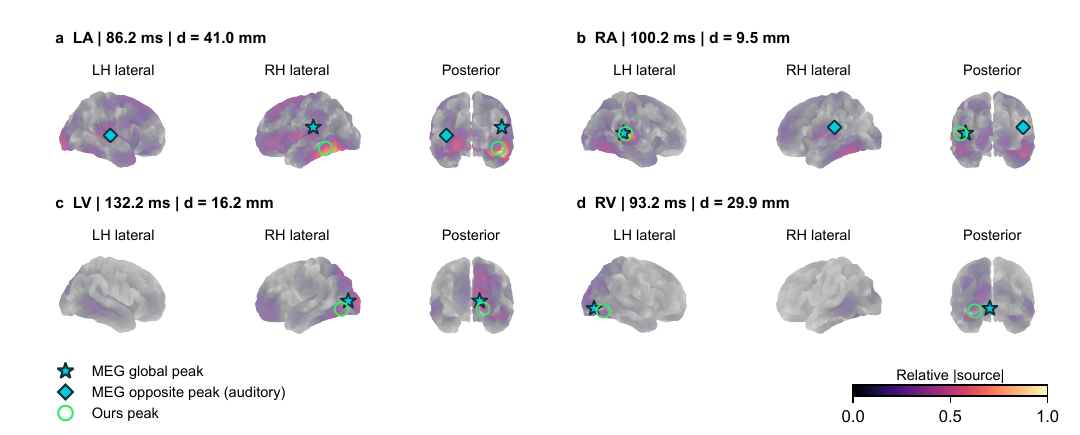}
\caption{MNE sample peak localization for (a) left ear, (b) right ear, (c) left visual field, and (d) right visual field. Each condition shows MEG-dSPM and Ours at the source-maximum time of Ours, in left-lateral, right-lateral, and posterior views. Cyan stars mark the global MEG peak, cyan diamonds the other-hemisphere auditory MEG peak, and green circles the peak of Ours. Distances $d$ are measured in canonical three-dimensional source coordinates, not in the displayed projections. Absolute maps are independently whole-brain peak-normalized, with shared linear magma color and square-root opacity and no threshold.}
\label{fig:vep}
\end{figure*}

Table~\ref{tab:mne_sample} summarizes the results, and Fig.~\ref{fig:vep} shows the corresponding maps. For both auditory conditions, the proposed method gave the smallest distance; the learned residual reduced the right-ear distance relative to ridge completion and did not change the left-ear peak. For visual stimulation, it did not improve on the closest baselines, MNE for the left field and dSPM for the right field. Overall, the proposed method showed the closest agreement with the MEG reference for auditory responses, whereas the conventional baselines yielded smaller peak distances for visual responses.

\begin{table}[t]
\caption{Global peak distances (mm) to MEG-dSPM at each method's own source-maximum time over 29.2--156.2~ms. Distances are computed from the selected source-maximum frame. Bold marks the smallest value among the listed methods, including ties.}
\label{tab:mne_sample}\centering\footnotesize
\setlength{\tabcolsep}{2.5pt}
\begin{tabular*}{\columnwidth}{@{\extracolsep{\fill}}lrrrr@{}}
\toprule
Method & Left ear & Right ear & Left field & Right field\\
\midrule
MNE & 57.88 & 45.63 & \textbf{11.80} & 22.60\\
dSPM & 47.49 & 36.45 & 38.00 & \textbf{16.56}\\
sLORETA & 62.84 & 80.51 & 15.92 & 16.71\\
Ridge & \textbf{41.03} & 27.54 & 22.79 & 29.87\\
Ours & \textbf{41.03} & \textbf{9.47} & 16.20 & 29.87\\
\bottomrule
\end{tabular*}
\end{table}

\subsection{Coreg-Spikes Evaluation}
\label{sec:coreg}
\begin{figure*}[t]
\centering
\includegraphics[width=0.9\textwidth,trim=0 0 0 12,clip]{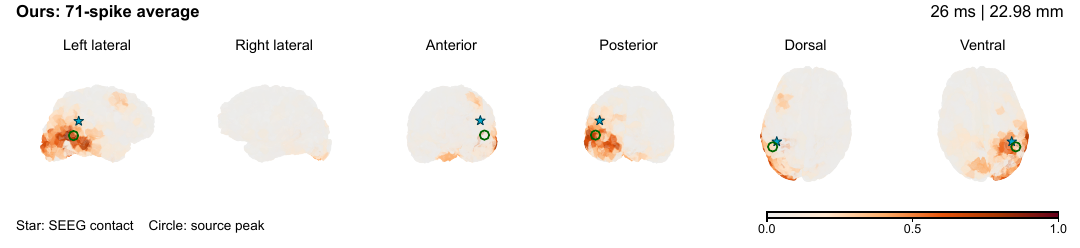}
\caption{Coreg-spikes sub-01 reconstruction from the average of 71 spike-aligned EEG epochs. Ours denotes the individual-geometry Posterior-DDIM row + ridge + TCN pipeline. Six views show the same peak-normalized absolute source map at 26~ms, the time of the space--time source maximum over $-10$ to 100~ms. Cyan stars mark the projected SEEG contact, and green circles mark the reconstruction peak; the two are 22.98~mm apart in three dimensions. All views share one normalization, and no threshold is applied. Projected markers need not lie on the visible surface.}
\label{fig:coreg_spikes}
\end{figure*}

We also evaluated participant sub-01 of the coreg-spikes dataset, which provides simultaneous scalp EEG and stereo-EEG (SEEG) \cite{Zauli2024CoregSpikes}, using an individual-geometry prior, ridge map, and TCN trained only on simulated sources for this forward model. The TCN reduced LE relative to ridge completion from 50.48 to 32.47~mm, so the completion remained effective once retrained for an individual forward model. For the average of 71 spike-aligned epochs, the space--time peak of the proposed method at 26~ms lay 22.98~mm from the SEEG contact, compared with 36.51~mm for MNE, 28.70~mm for dSPM, and 66.30~mm for sLORETA (Fig.~\ref{fig:coreg_spikes}).

\section{Discussion}
\label{sec:discussion}

A central feature of the proposed framework is the explicit separation of measurement-constrained and prior-driven information. The row-space component is refined using both the EEG and the learned diffusion prior, whereas Stage II introduces only structure that is invisible to the measurement operator. The matched ablations showed that posterior row refinement and nonlinear null-space completion contributed distinct improvements to localization when the other component was held fixed (Section~\ref{sec:ablation}). Because the completion is restricted to the lead-field null space, it cannot alter the EEG predicted by the refined row estimate. Improvements introduced by Stage II therefore arise from learned source structure rather than additional fitting of the measurements. At the same time, this separation makes the limitation of prior-driven completion explicit: the EEG provides no direct means to verify structure added in the null space, and inaccurate priors may introduce measurement-invisible activity \cite{Bhadra2021}.

Learned deterministic null-space completion improved support ranking relative to retaining the sampler’s null realization (Section~\ref{sec:solvercomp}), with higher AP and an unresolved difference in LE. The predictor uses learned row–null dependence to construct a single completion for localization. These results support this design for the evaluated task and complement the squared-error motivation in (Section~\ref{sec:formulation}). The observed gain reflects the combined choice of a learned conditional predictor and deterministic completion. Extending Stage II to model $p(N\mid R)$ would enable uncertainty quantification within the null space.

The remaining localization error is strongly influenced by Stage I. The exact-row diagnostic shows that substantially better completion is possible when the true observable component is available, indicating that row-estimation error propagates directly into Stage II. A related source of mismatch is that the residual model is trained on clean row components but evaluated on estimated ones. Training the completion model on Stage I outputs may reduce this discrepancy and improve robustness to row-estimation errors.

The two stages also have different dependencies on the forward model. The row--null decomposition, ridge map, and residual basis are defined by the lead field, making the null-space completion specific to a given forward model. The diffusion prior is independent of EEG measurements but remains tied to the cortical source space on which it is trained. Consequently, deployment on a new participant requires an appropriate individual forward model and refitting of the completion modules, while a different cortical source space additionally requires retraining the source prior. The coreg-spikes experiment demonstrates adaptation to an individual geometry through geometry-specific training (Section~\ref{sec:coreg}).

The present evidence therefore has a defined scope. The simulation benchmark used a single lead field with fixed source orientations, TVB-generated activity, and independent Gaussian noise, while the recorded-data experiments involved one participant from each dataset and relied on cross-modal or single-contact references rather than known cortical ground truth. These experiments demonstrate simulation performance and condition- or event-dependent concordance on recorded data, but not validated clinical localization. Broader evaluation across participants, forward-model perturbations, and datasets with known source locations, such as intracranial stimulation recorded with high-density EEG \cite{Mikulan2020LocalizeMI}, will be important for establishing robustness and clinical utility. Finally, the present study evaluates localization rather than absolute source amplitude, and the completed null component should be interpreted as a representative estimate supported by the learned source distribution rather than a uniquely recovered cortical waveform.

\section{Conclusion}

We introduced an observability-aligned formulation of EEG source localization that treats the observable and unobservable components of the inverse problem as distinct estimation tasks. Singular-coordinate Posterior-DDIM refines the noisy row-space component by balancing a learned diffusion prior against direction-dependent measurement reliability, while deterministic ridge and nonlinear residual models complete the null component from the refined row representation. Restricting the completion to the lead-field null space preserves the refined row estimate and its predicted EEG exactly, allowing measurement-constrained reconstruction and prior-driven completion to remain explicitly separated. Across simulations, the proposed framework achieved higher frame-level AUC and average precision and lower localization error than eight baselines at every tested noise level. Matched ablations further attributed complementary gains to posterior row refinement and learned null-space completion, while deterministic completion improved support identification relative to retaining sampled null-space realizations. Exploratory applications to evoked EEG and simultaneous scalp EEG--SEEG data showed condition- and event-dependent agreement with cross-modal references. These results support observability-guided diffusion reconstruction as a principled approach to severely underdetermined EEG source localization, while highlighting improved row estimation, robustness to model mismatch, and validation with known source locations as important directions for future work.

\section*{Acknowledgment}
Funding and non-anonymous acknowledgments will be inserted after review.

\balance
\bibliographystyle{IEEEtran}
\bibliography{references}

\end{document}